\documentclass[twocolumn]{article}

\usepackage{authblk}
\usepackage{hyperref}
\usepackage[margin=2cm]{geometry}
\usepackage{graphicx}
\usepackage{dcolumn}
\usepackage{bm}
\usepackage{physics}
\usepackage{authblk}

\newcommand{\R}{{\bf R}}

\providecommand{\keywords}[1]
{
  \small	
  \textbf{\textit{Keywords---}} #1
}

\begin{document}

\title{Enhanced electron-phonon coupling near an electronic quantum phase transition}
\author{Nikitas Gidopoulos \\
\small Department of Physics, Durham University, South Road, Durham, DH1 3LE, UK}%\footnote{nikitas.gidopoulos@durham.ac.uk}}
%\affil{\small Department of Physics, Durham University, South Road, Durham, DH1 3LE, UK}
\date{October 1, 2021}

\twocolumn[
  \begin{@twocolumnfalse}
    \maketitle
    	\begin{abstract}
	I construct a simple model to demonstrate that when the many-electron quantum state of a material is near a quantum phase transition and the vibrational motion of a phonon  explores the potential energy surface near the transition 	point, then an impenetrable barrier appears at the potential energy surface which restricts the phonon from crossing the transition point and abnormally increases the phonon frequency. 
	The ensuing anomalous enhancement of the electron phonon coupling is general and independent of the specific nature of the electronic quantum phase transition. Understanding and modelling this strong electron-phonon coupling may  potentially lead to the design of phonon superconductors with high critical temperatures by choosing their parameters appropriately near an electronic quantum phase transition.
	 
	\end{abstract}
 \\

  \end{@twocolumnfalse}
]

%\begin{multicols}{2}

\section{Introduction}

The coupling between electrons and phonons is investigated when a phonon drives an electronic system towards a transition of the many-electron state. To analyse this effect, it is useful to review briefly the adiabatic separation of the electronic and nuclear degrees of freedom\cite{bo,born_huang,handy1996adiabatic}. 

It is well known that the large difference between the masses and energy scales of electrons and nuclei allows the separate treatment of their respective motions. The electrons are so much faster, that for them, nuclear motion is a slow, adiabatic change. The corresponding electronic energy, dependent on the nuclear positions, provides the potential energy surface (PES) where the nuclei move. 
In periodic solids, as the nuclei slowly vibrate around their equilibrium positions, the electronic wave function evolves adiabatically with them. Traditionally, the evolved wave function is expanded in terms of the electronic eigenfunctions for the periodic structure and in the deviations of the phonons from equilibrium. The coefficients of the expansion give the well-known electron-phonon couplings\cite{giustino2017electron}.

In the following section, to investigate what happens when a phonon drives an electronic system towards a transition of the many-electron state, I construct a simple one-dimensional model where a phonon with normal coordinate $Q$ is coupled to an electronic state that depends parametrically on $Q$ and undergoes a transition at $Q=Q_0$.

%\section{Model of phonon overlapping quantum critical point} % undergoing transition}
\section{Model of phonon overlapping electronic transition}

%To construct our simple model, 
To model the $Q$-dependent electronic state with a transition at $Q_0$, I employ the ground state of the eigenvalue problem:
\begin{equation} 
\label{eigv}
A \left[
\begin{array}{cc}
k (Q - Q_0) & \alpha \\
\alpha & k ( Q_0 - Q )  
\end{array}
\right] \Big( { a_Q \atop b_Q} \Big) = \epsilon_Q \, \Big( { a_Q \atop b_Q} \Big) 
%+ \frac{k^2}{2} + D
\end{equation}
The parameter $k$ is the strength of the (linear) electron-phonon coupling. 
The small parameter $\alpha > 0$ controls the speed of transition of the electronic ground state 
between the two states $\Big( {\displaystyle{ 1 \atop 0}} \Big) $, $\Big( {\displaystyle{ 0 \atop 1}} \Big) $. 
When $\alpha \rightarrow 0$, the transition becomes abrupt and we can employ the ground state of \eqref{eigv}
%$\bigg( {\displaystyle{ a_Q^{-} \atop b_Q^{-}}} \bigg) $ 
to model a quantum phase transition (QPT)\cite{sachdev2011quantum} %of the electronic state 
between two qualitatively different many-electron states. 
On the other hand, for finite $\alpha $ we have a gradual transition, or a cross-over region connecting the two electronic states on either side of $Q_0$. 

In order to model an electronic QPT at $Q_0$ without a level crossing\footnote{To avoid couplings to higher electronic levels and allow the model to be described accurately in the adiabatic approximation.}, I multiply the $2 \times 2$ matrix in \eqref{eigv} by $A=1+D/\lambda_Q$, where $D \gg 1$ is an energy scale much larger than the harmonic phonon frequency and $\lambda_Q= \sqrt{k^2 (Q-Q_0)^2+\alpha^2}$.
%
%The superscripts $-$, $+$, denote the electronic ground, excited states. 
The energy eigenvalues in \eqref{eigv} are $\epsilon_Q^{\pm} = \pm (D+ \lambda_Q)$. 
%$\lambda_Q^{\pm} = \pm \sqrt{k^2 (Q-Q_0)^2+\alpha^2}$.
%In Fig.~\ref{fig:epsilon_Q} the lowest eigenvalue $\lambda_Q^-$ is shown for different values of $\alpha$. 
%The model mimics an electronic transition in the limit  $\alpha \rightarrow 0$. 
%\begin{figure}%[tbhp]
%\centering
%\includegraphics[width=.6\linewidth]{epsilon_Q.pdf}
%\caption{The electronic ground state energy for various values of $\alpha$. 
%The model shows an electronic quantum phase transition for $\alpha \rightarrow 0$.}
%\label{fig:epsilon_Q}
%\end{figure}
%
%
%Using similar notation, t
The electronic ground/excited states are %denoted by 
$\bigg( {\displaystyle{ a_Q^{\pm} \atop b_Q^{\pm}}} \bigg) $ where,
\begin{eqnarray}
a_Q^{\pm} &=& \pm \frac{1}{\sqrt{2}} \sqrt{1 \pm \frac{Q-Q_0}{\sqrt{( Q-Q_0)^2+(\alpha/k)^2}} } , \label{aq} \\
b_Q^{\pm} &=&  \frac{1}{\sqrt{2}} \sqrt{1 \mp \frac{Q-Q_0}{\sqrt{(Q-Q_0)^2+(\alpha/k)^2}} } \,  . \label{bq}
\end{eqnarray}
%As $Q$ varies, the electronic ground state changes:
%\begin{equation}
%\bigg( { a_Q^{-} \atop b_Q^{-}} \bigg)  = \left\{ 
%\begin{array}{ccl}
%- \Big( {\displaystyle {1 \atop 0}} \Big) &{\rm for} & Q \ll Q_0 , \\ 
%\\
%{\displaystyle \frac{1}{\sqrt{2}} \Big( { -1 \atop 1} \Big) } & {\rm at } & Q=Q_0 , \\
% \\
%{\displaystyle  \Big( {0 \atop 1} \Big) } &  {\rm for} & Q \gg Q_0 .
%\end{array}
%\right.
%\end{equation}
%
%

To proceed with the construction of the model,
%with a phonon coupled with the electronic ground state. 
%$\bigg( {\displaystyle{ a_Q^{-} \atop b_Q^{-}}} \bigg) $. 
I choose (for zero electronic coupling, $k=0$) a harmonic phonon, with unit mass and oscillator strength. The harmonic oscillator potential has its minimum at $Q=0$, slightly offset from the QPT at $Q_0 > 0$.
The Hamiltonian of the combined electron-phonon system is (in atomic units, $\hbar = 1$):
\begin{equation} \label{H_e_ph}
H = 
-\frac{1}{2} \frac{d^2}{dQ^2} + \frac{Q^2}{2} \\
+ A
\left[
\begin{array}{cc}
k(Q - Q_0) & \alpha \\
\alpha & k(Q_0 - Q) 
\end{array}
\right] 
%+ \frac{k^2}{2} + D
\end{equation}

Because a large gap of magnitude, $\epsilon_{Q_0}^{+} - \epsilon_{Q_0}^{-} = 2 (D+\alpha)$, separates the two adiabatic electronic energy levels, 
the electron-phonon states $| \Psi_n ( Q ) \rangle $ are described accurately in the adiabatic approximation\cite{born_huang,handy1996adiabatic}.  
Hence, we can write for the electron-phonon states $| \Psi_n ( Q ) \rangle$ involving the adiabatic electronic ground state, 
\begin{equation} \label{eph_state}
| \Psi_n ( Q ) \rangle \simeq \bigg( { a_Q^{-} \atop b_Q^{-}} \bigg) \, X_n (Q)  .
\end{equation}
%$X_n (Q)$ is the adiabatic phonon wavefunction in the adiabatic electronic ground state PES.
%
The adiabatic phonon wavefunctions $X_n (Q)$ satisfy the one-dimensional 
adiabatic phonon Schr\"odinger's equation,
\begin{equation} \label{Xeq}
\left[ -\frac{1}{2} \frac{d^2}{dQ^2} + V(Q) \right] X_n(Q) = E_n \, X_n (Q) ,
\end{equation}
where $V(Q)$ is the (anharmonic) PES corresponding to the adiabatic electronic ground state
\begin{equation} \label{V}
V(Q) =  \frac{1}{2} Q^2 - \sqrt{k^2 (Q-Q_0)^2+\alpha^2} . %+ V (Q) , 
\end{equation}
In Eqs.~(\ref{Xeq},\ref{V}) I have shifted the PES and phonon energy by the constant $D$. 

\begin{figure}[h]
\centering
\includegraphics[width=8.0cm]{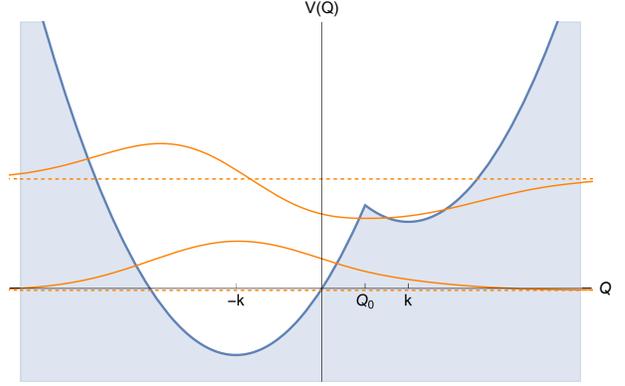}
\caption{Anharmonic PES (blue) for $\alpha \rightarrow 0$, omitting the DBOC, showing the 
phonon ground and first excited state wavefunctions (solid lines) and energies (dashed lines) in orange.}
\label{fig:VQwf1}
\end{figure}

The PES where the phonon vibrates is shown in Fig.~\ref{fig:VQwf1}. 
%in blue for $\alpha \rightarrow 0$. 
When $\alpha = 0$, the model mimics a QPT of the adiabatic electronic ground state at $Q_0$, at the cusp in the PES.

\section{Strong electron-phonon coupling near a QPT}

I now investigate if anything dramatic happens to the coupling between the electronic state undergoing transition and the phonon when we solve the adiabatic phonon Schr\"odinger \eqref{Xeq} (using $k=1$, $Q_0=0.5$ in a.u.).  
I used Mathematica\cite{Mathematica} to solve numerically \eqref{Xeq} for the phonon ground and first excited state wavefunctions (solid lines) and energies (dashed lines), which are shown in Fig.~\ref{fig:VQwf1} in orange. 
The energy gap between the two energy levels gives an estimate for the phonon frequency. 
(The PES is anharmonic and the low lying phonon energy levels are not equally spaced.)

Having chosen the parameters of the model appropriately, the wavefunction for the vibrating phonon coordinate $Q$ overlaps the QPT at $Q_0$, Fig.~\ref{fig:VQwf1}. Hence, the zero-point motion allows the phonon to explore the neighbourhood of the QPT. 
In doing so, it carries with it, adiabatically, the electronic wavefunction \eqref{eph_state}. Consequently, according to the solution of this model shown in Fig.~\ref{fig:VQwf1}, the material is predicted to undergo successive phase transitions (at the phonon zero-point frequency) as the phonon oscillates crossing the QPT at $Q_0$.
%, or perhaps to be in a superposition of the many-electron states on either side of $Q_0$. 
However, this result is unphysical; such recurring phase transitions are not observed in real materials.
%, nor the superposition of macroscopically different states has been observed in real materials (but see Ref.~\onlinecite{freedman}).

One might expect that non-adiabatic effects\cite{Gidopoulos2005,gidopoulos2014electronic} would come into play and change the adiabatic picture\cite{handy1996adiabatic}.
However, in our model non-adiabatic effects are small (by choosing $D \gg 1$). 
In any case, the unphysical prediction remains valid for materials well described by the adiabatic approximation.

The paradox is resolved by including a term that is routinely omitted in the adiabatic approximation for solids, 
but in this case it becomes important. 
%we have neglected so far in the PES. 
%
In the adiabatic approximation, the potential surface includes a correction energy term, known in the literature as the diagonal Born-Oppenheimer correction (DBOC)\cite{handy1996adiabatic,gauss2006analytic,schneider2019diagonal}. The correction term is given by,
\begin{multline}
\label{ac}
W ( {\R_1, \ldots , \R_N} ) = \\
\sum_a \frac{\hbar^2}{2 M_a} \langle \nabla_{\R_a} \Phi_{\R_1, \ldots , \R_N} | \nabla_{\R_a} \Phi_{\R_1, \ldots , \R_N} \rangle      
\end{multline} 
where $| \Phi_{\R_1, \ldots , \R_N} \rangle $ is the many-electron ground state. In the adiabatic approximation the state 
$| \Phi_{\R_1, \ldots , \R_N} \rangle $ depends parametrically on the nuclear positions, $\{ \R_1, \ldots , \R_N \}$. 
The sum is over the nuclei, $1 \le a \le N$, the gradients $\nabla_{\R_a} $ are over the nuclear positions; $M_a$ 
are the nuclear masses and the bra-ket notation shows integration over electronic degrees of freedom.
A simple inspection of the term reveals that $W ({\R_1, \ldots , \R_N})$ 
is a positive (repulsive) energy term and that it is a measure of the rate of change of the electronic state 
$| \Phi_{\R_1, \ldots , \R_N} \rangle $ when the nuclear positions change, e.g. along a phonon. 
Because each term of the sum is divided by the corresponding nuclear mass, the DBOC term is thought to be vanishingly small and is typically omitted. 
However, this omission is not justified when a phonon drives the many-electron ground state towards a transition: 
The rapid change of the electronic wave function in the vicinity of the transition makes the derivatives of electronic state grow and ultimately diverge at the point of the transition. 
Division by the large nuclear mass can no longer give a vanishingly small term. As a result, the DBOC energy term increases and takes the shape of 
a barrier in the PES, preventing the phonon from crossing the transition point and thereby strongly affecting its energy and wave function.

In the following, I illustrate this effect with the help of our model. 
We shall see that inclusion of the DBOC leads to an abnormal increase of the phonon frequency. % near an electronic quantum phase transition.

The DBOC in \eqref{ac} becomes for our model, 
\begin{equation} \label{dbv0}
W(Q) = \frac{1}{2} \left( \frac{d a_Q^{-} }{d Q} \right)^2 + \frac{1}{2} \left( \frac{d b_Q^{-} }{d Q} \right)^2 .
\end{equation}
From the $Q$-dependent electronic state $\bigg( {\displaystyle{ a_Q^{-} \atop b_Q^{-} }} \bigg) $ we can obtain the DBOC energy term. % (Fig.~\ref{fig:dboc}). 
Using Eqs.~(\ref{aq},\ref{bq},\ref{dbv0}) we obtain
\begin{equation}
W(Q) = \frac{1}{8} \, \frac{ (\alpha/k)^2 }{ \big[(Q-Q_0)^2+(\alpha/k)^2 \big]^2}  .
\end{equation}
$W(Q)$ is a function peaked at $Q_0$. In the limit $\alpha \rightarrow 0$, it %$W(Q)$ 
becomes a delta function barrier with diverging weight, proportional to $1/\alpha$.
%\begin{figure}%[tbhp]
%\centering
%\includegraphics[width=.8\linewidth]{dboc.pdf}
%\caption{The DBOC as a function of $Q$ for various values of $\alpha$. 
%The DBOC becomes a delta function with diverging weight inversely proportional to 
%$\alpha$ for $\alpha \rightarrow 0$.}
%\label{fig:dboc}
%\end{figure}
Scattering of a wave on a one-dimensional delta function barrier with infinite weight gives transmission coefficient equal to zero and reflection coefficient equal to one.

%$$***$$

\begin{figure*}[t]
\centering
\includegraphics[width=8.0cm]{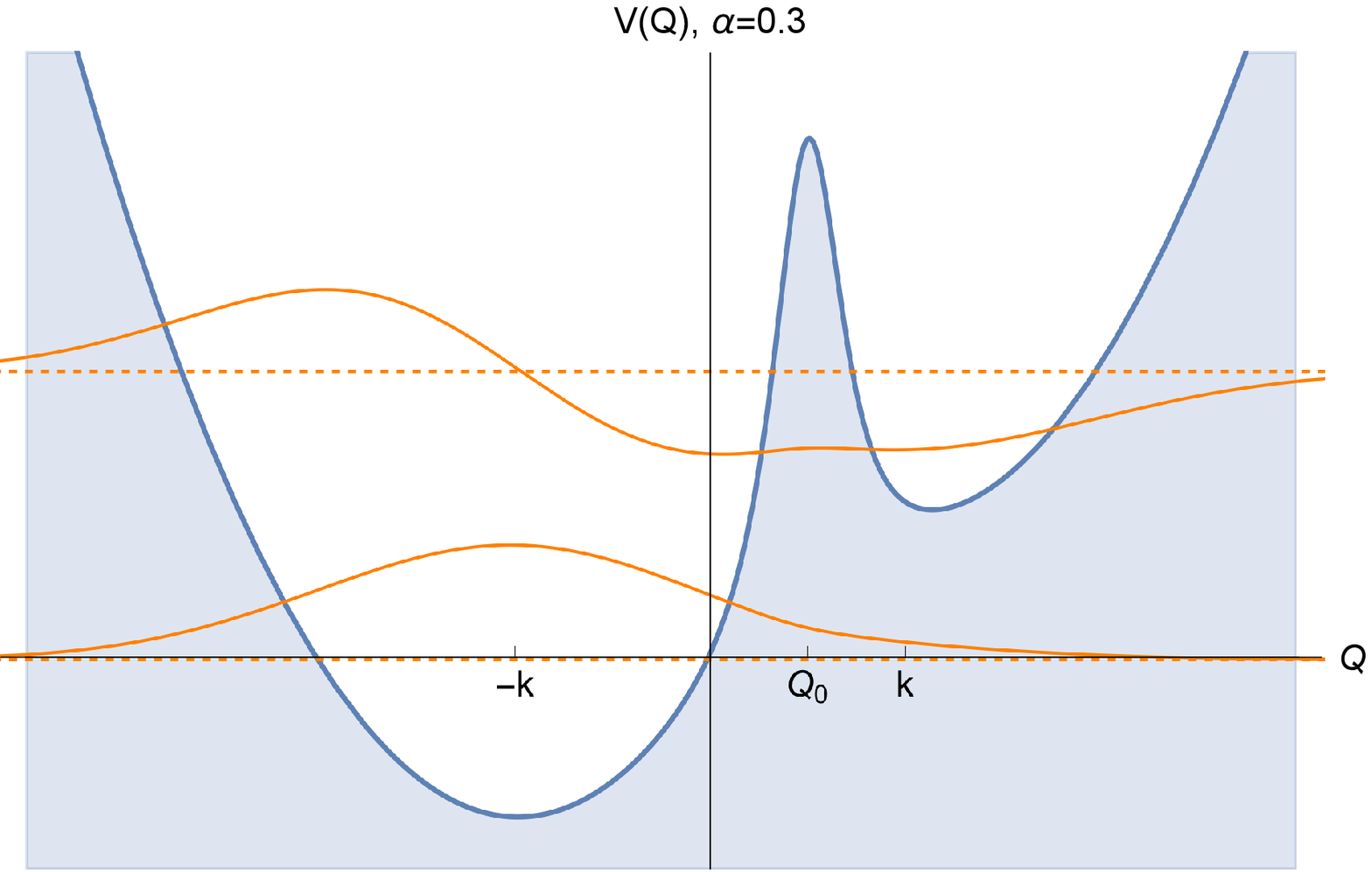}
\includegraphics[width=8.0cm]{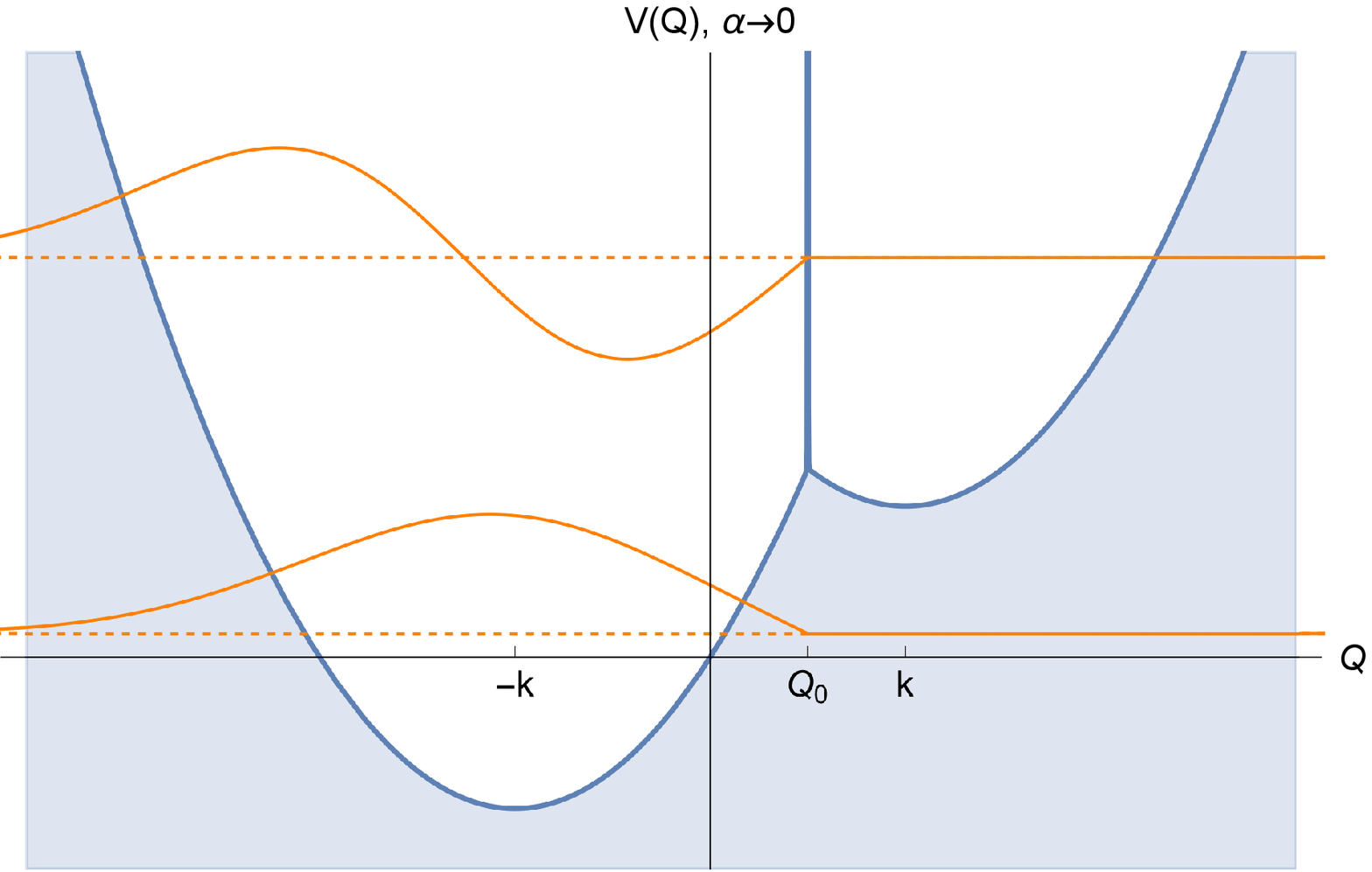}
\caption{Blue: PES including the DBOC for two different $\alpha $. When $\alpha \rightarrow 0$ the DBOC has become an impenetrable delta function barrier. Orange: ground and first excited state phonon wavefunctions (solid lines) and energies (dashed lines).}
\label{fig:3delta}
\end{figure*}

Fig.~\ref{fig:3delta} shows the potential energy surfaces, including the DBOC barrier for a small $\alpha$ and for $\alpha \rightarrow 0$, together with the phonon wavefunctions and energies. 
It is evident, that introducing the DBOC for a finite $\alpha$, the phonon wavefunctions %tend to be confined to the left of the transition and 
tunnel less to the right side of the transition. At the same time, the phonon frequency (gap) increases.
In the limit $\alpha \rightarrow 0$, the DBOC term has become an infinitely thin but impenetrable, delta-function barrier, confining perfectly the phonon wave functions 
on the left of the transition point. 

% \begin{SCfigure*}[\sidecaptionrelwidth][t]
% \centering
% \includegraphics[width=5.9cm]{V_Q_dd_0dot3_wf.pdf}
% \includegraphics[width=5.9cm]{V_Q_dd_0dot1_wf.pdf}
% \includegraphics[width=5.9cm]{V_Q_dd_0dot01_wf.pdf}
% \caption{Blue: PES including the DBOC for various $\alpha $. 
% Orange: ground and first excited state phonon wavefunctions (solid lines) and energies (dashed lines).}
% \label{fig:2delta1}
% \end{SCfigure*}

% \begin{SCfigure*}[\sidecaptionrelwidth][t]
% \centering
% \includegraphics[width=5.7cm]{V_Q_dd_0dot01_wf.pdf}
% \includegraphics[width=5.7cm]{V_Q_dd_0dot00001_wf.pdf}
% \caption{Blue: PES including the DBOC for small $\alpha $. When $\alpha \rightarrow 0$ the 
% DBOC has become an impenetrable delta function barrier. 
% Orange: ground and first excited state phonon wavefunctions (solid lines) and energies (dashed lines).}
% \label{fig:3delta2}
% \end{SCfigure*}

%\begin{figure}%[tbhp]
%\centering
%\includegraphics[width=.8\linewidth]{V_Q_dd_0dot00001_wf.pdf}
%\caption{Blue: PES including the DBOC for small $\alpha $. 
%When $\alpha \rightarrow 0$ the 
%DBOC has become an impenetrable delta function barrier. 
%Orange: ground and first excited state phonon wavefunctions (solid lines) and energies 
%(dashed lines).}
%\label{fig:3delta2}
%\end{figure}

This picture is physically intuitive, especially compared with the prediction when the DBOC is ignored, that phonons near e.g. a metal to Mott-insulator transition can tunnel between the two regions and thus induce recurring metal-insulator transitions, or perhaps bring the electronic state in a superposition of entirely different macroscopic states (but see Ref.\cite{friedman2000quantum}).
%, which is the prediction when the DBOC is ignored.} 
%Nevertheless, without the DBOC barrier in the PES, there is nothing in the adiabatic equations to prevent phonons from doing just that.}

\begin{table}[tbhp]
\centering
\setlength\tabcolsep{3pt}
{\bf Increase of phonon frequency with DBOC}
\begin{tabular}{cccccc}
&$\alpha = 0.3 $ & $\alpha = 0.2$ &  $\alpha = 0.1$ & $\alpha = 0.01$&  $\alpha \rightarrow 0 $\\
\hline
%$\hbar \omega$  & 0.953509 & 1.02722 & 1.13007 & 1.23627 & 1.24439 \\
$\omega$  & 0.954 & 1.027 & 1.130 & 1.236 & 1.244 \\
%$\hbar \Delta \omega$  & 0.11712 & 0.19083 & 0.29368 & 0.39988 & 0.408\\
%$\Delta \omega / \omega$ & 0.14003 & 0.22815 & 0.35112 & 0.4781 & 0.48781 \\
$\Delta \omega / \omega_0$ & 0.14 & 0.23 & 0.35 & 0.48 & 0.49 \\
\hline
\end{tabular}
\caption{Phonon frequencies in atomic units, approximated as the energy differences between the two lowest phonon energy levels, when the DBOC is included. Smaller value of $\alpha$ indicates a faster transition. The difference, $\Delta \omega = \omega - \omega_0$, is from the reference value $\omega_0 = 0.836$ au, when the DBOC is omitted, see Fig.\ref{fig:VQwf1}.}
\label{tab:tt}
\end{table}

Finally, Table~\ref{tab:tt} shows the increase and relative increase of the phonon frequency, due to the confinement from the DBOC barrier. The jump for $\alpha \rightarrow 0$ is almost 50\%. I note the critical temperature in phonon-mediated superconductivity increases quasi-linearly with phonon frequency\cite{pickett2008next}.
%\begin{figure}%[tbhp]
%\centering
%\includegraphics[width=.8\linewidth]{dw_over_w1}
%\caption{The relative increase of the phonon frequency for decreasing $\alpha$.}
%\label{fig:dw_over_w1}
%\end{figure}

\section{Discussion}

For materials described accurately by the adiabatic approximation, when a phonon brings the adiabatic electronic state near a QPT, a dramatic enhancement is predicted in the coupling between the phonon and the many-electron ground state.   
The coupling manifests primarily by the erection of a DBOC barrier that pushes the phonon wave function away from the transition thus raising abruptly the phonon frequency. 
The topology of the phonons is important, as the confinement of the phonon wave function is more effective in reduced dimensions.
The predicted mechanism is strongest near a true phase transition of the many-electron ground state but is still nonzero with a more gradual change between the two end-ground-states of the infinite system, as for example in the cross-over region between two electronic phases. In that case, a secondary effect is that since the PES changes with the mass-dependent DBOC term, then the dynamical matrix (Hessian) changes; the diagonalization of the corrected Hessian will give corrected normal modes, corrected phonon frequencies and corrected electron-phonon coupling matrix elements. The theoretical derivation of these corrections will be presented in a future publication.

In this work, the theoretical prediction 
%on the phonon frequencies and the confinement of phonon wave functions away from QPTs 
relies on the validity of the adiabatic approximation. In a forthcoming paper, I study the extension of the proposed mechanism to materials where the adiabatic approximation breaks down \cite{Gidopoulos2005,gidopoulos2014electronic}.

The immediate area of application of the theoretical prediction is in phonon-mediated superconductivity. 
In real materials, it is feasible to estimate quantitatively the DBOC (from \eqref{ac}) using ab-initio calculations. 
%It would be a significant advance if these calculations managed to discover strong DBOC barriers near electronic transitions, both for the deeper understanding we shall gain on the coupling between electrons and phonons and for the potential breakthroughs this new mechanism will bring in the design of a new generation of phonon superconductors at higher critical temperatures.
%
A possible theoretical discovery of strong DBOC barriers near electronic QPTs with such calculations would confirm the predicted mechanism and deepen our quantitative understanding of the coupling between electrons and phonons. 
It would also open the way to control the critical temperature of superconducting materials with a QPT, by tuning their parameters to bring the many-electron ground state optimally in the neighbourhood of the phase transition.

\section*{Acknowledgements}
I thank Prof. S.J. Clark and Prof. D. Hampshire for useful discussions.
%\end{acknowledgements}

% Bibliography
%\bibliographystyle{plain}
%\bibliography{pnas-sample}

\begin{thebibliography}{10}

\bibitem{bo}
M.~Born and R.~Oppenheimer.
\newblock Zur quantentheorie der molekeln.
\newblock {\em Annalen der Physik}, 389(20):457--484, 1927.

\bibitem{born_huang}
Max Born and Kun Huang.
\newblock {\em Dynamical theory of crystal lattices}.
\newblock Clarendon Press, 1968.

\bibitem{friedman2000quantum}
Jonathan~R Friedman, Vijay Patel, Wei Chen, SK~Tolpygo, and James~E Lukens.
\newblock Quantum superposition of distinct macroscopic states.
\newblock {\em Nature}, 406(6791):43--46, 2000.

\bibitem{gauss2006analytic}
J{\"u}rgen Gauss, Attila Tajti, Mih{\'a}ly K{\'a}llay, John~F Stanton, and
  P{\'e}ter~G Szalay.
\newblock Analytic calculation of the diagonal born-oppenheimer correction
  within configuration-interaction and coupled-cluster theory.
\newblock {\em The Journal of chemical physics}, 125(14):144111, 2006.

\bibitem{Gidopoulos2005}
N.~Gidopoulos and E.~Gross.
\newblock Electronic non-adiabatic states.
\newblock {\em arXiv:cond-mat/0502433}, 2005.

\bibitem{gidopoulos2014electronic}
Nikitas~I Gidopoulos and EKU Gross.
\newblock Electronic non-adiabatic states: towards a density functional theory
  beyond the born--oppenheimer approximation.
\newblock {\em Philosophical Transactions of the Royal Society A: Mathematical,
  Physical and Engineering Sciences}, 372(2011):20130059, 2014.

\bibitem{giustino2017electron}
Feliciano Giustino.
\newblock Electron-phonon interactions from first principles.
\newblock {\em Reviews of Modern Physics}, 89(1):015003, 2017.

\bibitem{handy1996adiabatic}
Nicholas~C Handy and Aaron~M Lee.
\newblock The adiabatic approximation.
\newblock {\em Chemical physics letters}, 252(5-6):425--430, 1996.

\bibitem{Mathematica}
Wolfram~Research{,} Inc.
\newblock Mathematica, {V}ersion 12.3.1.
\newblock Champaign, IL, 2021.

\bibitem{pickett2008next}
Warren~E Pickett.
\newblock The next breakthrough in phonon-mediated superconductivity.
\newblock {\em Physica C: Superconductivity}, 468(2):126--135, 2008.

\bibitem{sachdev2011quantum}
Subir Sachdev.
\newblock {\em Quantum phase transitions}.
\newblock Cambridge university press, 2011.

\bibitem{schneider2019diagonal}
Patrick~E Schneider, Fabijan Pavo\v{s}evi\v{c}, and Sharon Hammes-Schiffer.
\newblock Diagonal born--oppenheimer corrections within the nuclear--electronic
  orbital framework.
\newblock {\em The journal of physical chemistry letters}, 10(16):4639--4643,
  2019.

\end{thebibliography}

%\end{document}

%\end{multicols}
\end{document}